\documentclass{llncs}
\usepackage{graphicx}
\usepackage{hyperref}
\usepackage{xcolor}
\usepackage{listings}
\usepackage{fancyvrb}
\usepackage{balance}
\newcommand{\reduceparspace}[1]{\vspace{0mm}}
\usepackage{layouts}
\titlerunning{Context-aware Failure-oblivious Computing}

\begin{document}
\title{Context-aware Failure-oblivious Computing as a Means of Preventing Buffer Overflows\thanks{The final authenticated version is available online at \url{https://doi.org/10.1007/978-3-030-02744-5_29}. We thank Oracle Labs for funding this research. We thank Gerg\"o Barany, Roland Yap, and Fabio Niephaus for their useful feedback on an early draft of this paper. We thank Ingrid Abfalter for proofreading and editorial assistance.}}
\author{Manuel Rigger\inst{1} \and
Daniel Pekarek\inst{1} \and
Hanspeter M\"ossenb\"ock\inst{1}}
\authorrunning{M. Rigger et al.}
\institute{Johannes Kepler University Linz, Austria\\
\email{\{manuel.rigger,daniel.pekarek,hanspeter.moessenboeck\}@jku.at}}
\maketitle
\begin{abstract}
In languages like C, buffer overflows are widespread.
A common mitigation technique is to use tools that detect them during execution and abort the program to prevent data leakage or the diversion of control flow.
However, for server applications, it would be desirable to prevent such errors while maintaining availability of the system.
To this end, we present an approach to handling buffer overflows without aborting the program.
This approach involves implementing a recovery logic in library functions based on an introspection function that allows querying the size of a buffer.
We demonstrate that introspection can be implemented in popular bug-finding and bug-mitigation tools such as LLVM's AddressSanitizer, SoftBound, and Intel-MPX-based bounds checking.
We evaluated our approach in a case study of real-world bugs and show that for tools that explicitly track bounds data, introspection results in a low performance overhead.
\end{abstract}
\section{Introduction}

Buffer overflows in C, where an out-of-bounds pointer is dereferenced, belong to the most dangerous software errors~\cite{bufferoverflows,sok}.
Unlike higher-level languages, buffer overflows invoke \emph{Undefined Behavior} and are not prevented during execution; programmers also cannot handle them using exception or similar mechanisms, since the language lacks them.
Buffer overflows allow attackers to overflow function addresses stored on the stack or heap and thus to maliciously divert execution of the program~\cite{Shacham2007} and to leak sensitive data~\cite{Strackx2009}.
A plethora of tools exist that make their exploitation more difficult or detect them and abort execution of the program~\cite{memoryerrors,sok,runtime,Song2018}.
However, when availability of an application is important (e.g. for production servers), it would be preferable to continue execution as long as security is not compromised~\cite{acceptabilitycomputing}.
This could, for example, make it harder to perform a denial-of-service attack where a buffer overflow is exploited to crash the program or inject code.

To safely maintain execution in the presence of buffer overflows, we have come up with the concept of \emph{context-aware failure-oblivious computing}.
Our core idea is that library writers (e.g., the libc maintainers) can query run-time data such as bounds information in library functions by using an introspection interface.
This information can then be used to implement a recovery logic that can mitigate incorrect execution states instead of aborting the program.
Library writers can implement a custom recovery logic that depends on each function's semantics, which is why we refer to our technique as being context-aware.
For example, a libc function that processes an unterminated string could prevent an out-of-bounds access by checking for the end of the buffer to handle the fault and continue execution.
We expect that this recovery logic would be used mainly in a production context, as it would be preferable that execution is aborted if an error occurs during development and testing so that programmers can fix the error. 

\sloppy
Our work is based on a combination of \emph{failure-oblivious computing}~\cite{failureoblivious} and our previous work on an introspection interface for C to increase the robustness of libraries~\cite{introspection}.
We show how the introspection interface can be used to implement a failure-oblivious computing mechanism.
We evaluated our approach by demonstrating that introspection for preventing buffer overflows can be implemented in popular bug-finding and bug-mitigation tools such as LLVM's AddressSanitizer~\cite{asan}, SoftBound~\cite{softbound}, and GCC's Pointer Bounds Checker, which is based on the Intel Memory Protection Extensions (MPX)~\cite{mpxexplained}.
Furthermore, we show how our approach allows execution to continue in the presence of buffer overflows found in real-world programs as described by the Common Vulnerabilities and Exposures (CVE) database~\cite{cve}, and demonstrate that the performance overhead for introspection implemented in approaches such as MPX is negligible.

\section{Background}

\paragraph{Failure-oblivious computing.}
One technique for maintaining availability in the presence of buffer overflows is \emph{failure-oblivious computing}, where invalid writes are discarded and values for invalid reads are manufactured~\cite{failureoblivious,boundlessmemoryblock}.
By carefully selecting a sequence of return values for invalid reads, the program can successfully continue execution in most cases.
However, a drawback of this approach is that it is ``blind''; that is, it cannot guess the context (i.e., a function's semantics) to return a meaningful value for all reads.
In this paper, we address this aspect by making failure-oblivious computing context-aware.
\reduceparspace{}
\paragraph{Introspection for C.}
As part of previous work, we demonstrated how use of introspection (i.e., exposing run-time data) benefits the robustness of libraries~\cite{introspection}.
The core idea of our approach was that bug-finding tools and runtimes for C that track additional metadata such as object bounds or object types can expose this data to library writers via an introspection interface, which programmers can use to check the input of library functions.
We showed that various introspection functions can be used to detect bugs or to maintain availability of the program.
For example, to detect buffer overflows by means of introspection, the {\tt \_size\_right()}
\noindent function can be applied, which expects a pointer and returns the number of allocated bytes to the right of the pointee (or zero for invalid pointers) and can therefore be used for bounds checks.
In this paper, we expand on how introspection can be used to increase availability, which we define as \emph{context-aware failure-oblivious computing}.

\reduceparspace{}
\paragraph{Evaluation of introspection.} We have previously evaluated an introspection libc using Safe Sulong~\cite{lenientc,asplos}, an LLVM IR interpreter on top of the Java Virtual Machine (JVM)~\cite{native-sulong} which automatically keeps track of array lengths, object sizes, and object types of C data~\cite{introspection}.
Although the JVM tracks all relevant run-time information necessary to implement our introspection mechanism, it is not a typical environment in which to execute C code.
In this paper, we address this by evaluating our approach in the context of popular bug-finding and bug-mitigation tools for buffer overflows and show that our refined introspection approach prevents real-world errors while maintaining availability.

\section{Introspection Interceptors}
This section explains the implementation of the introspection-based libc functions.
These enhanced functions rely on the {\tt \_size\_right()} introspection function to mitigate buffer overflows.
Challenges to introducing them were that the original code not be cluttered by the introspection checks, that the effort for implementing these checks be low, and that the code behave in the same way as the original library during correct execution.

\paragraph{Libc interceptors.}
Based on our requirements, we implemented the introspection-based libc functions as \emph{interceptors}, which are wrappers that intercept calls to libc functions and which are used by many bug-checking and bug-mitigation tools (including ASan, GCC's Pointer Bounds Checker and SoftBound)\footnote{Note that in our previous work we instead reimplemented parts of a libc to use introspection, which made the libc less readable and required programs to be compatible with this libc.}.
The introspection logic was kept separate from the normal code to avoid cluttering of the original source code.
The cost of adding introspection-based recovery logic was low, as for each unsafe function that we considered (e.g., {\tt strlen()}), libc provides safer functions that expect an additional size argument, which we used for our implementation (e.g., {\tt strnlen()}).
By reusing existing libc functions from the same library, we expect correct execution to behave in the same way as without the interceptors.
For example, consider our {\tt strlen()} interceptor, which is based on the safer {\tt strnlen()} function:
{\begin{small}
\begin{verbatim}
size_t strlen(const char *s) {
  return ORIGINAL(strnlen)(s, _size_right(s));
}
\end{verbatim}
\end{small}
}
\noindent 
The {\tt ORIGINAL} macro yields a reference to the function passed as its argument that is part of the library and prevents recursively calling interceptors.
We implemented the {\tt \_size\_right()} function in various memory-safety-checking tools, as described in Section~\ref{sec:tools}.
Both the original {\tt strlen()} implementation and this interceptor behave correctly for strings that are terminated with a `\textbackslash{}0', which is needed to determine their length.
However, if an unterminated string is passed to the original {\tt strlen()} implementation, the function results in a buffer overflow that causes bug-finding and bug-mitigation tools to abort execution.
Using the introspection-based interceptor instead prevents the buffer overflow, as the string length can be computed even for strings for which the `\textbackslash{}0' is missing, because the interceptor assumes the underlying buffer size to be the maximum length of the string.
Note that application-level functions can still cause bug-finding and bug-mitigation tools to abort execution if these functions run over string bounds.
However, in many cases, application-level functions process strings up to the length computed by {\tt strlen()}, which consequently prevents an out-of-bounds access. 

As another example, an introspection interceptor can address the insecure interface of {\tt gets()}, which reads user input and writes it to a buffer whose size is unknown to the function:
{\begin{small}
\begin{verbatim}
char *gets(char *s) {
  return ORIGINAL(fgets)(s, _size_right(s), stdin);
}
\end{verbatim}
\end{small}
}
\noindent Using introspection, {\tt gets()} reads only as much user input as the buffer can store.

Some introspection interceptors correct invalid parameters, for instance, in {\tt memcpy}:
{\begin{small}
\begin{verbatim}
void *memcpy(void *dest, const void *src, 
             size_t n) {
  ssize_t dstsz = _size_right(dest);
  size_t len = n;
  if (dstsz < len) {
    len = dstsz;
  }
  return ORIGINAL(memcpy)(dest, src, len);
}
\end{verbatim}
\end{small}
}
\noindent If the size of the destination buffer is smaller than the number of bytes that the function is expected to copy, the function ignores the writes that go out of bounds.
Note that another check for the size of the source buffer would be applicable.

In contrast to our previous work~\cite{introspection}, we treat the return value of {\tt \_size\_right()} as a conservative estimate of the object's right bounds.
This estimate can be the real size of the object, in which case the introspection interceptors work most reliably.
However, it can also be at least as large as the actual allocation, which could include additional space due to alignment requirements (e.g., to accommodate approaches that track only allocation sizes).
Finally, if no bounds information is available for a given pointer, returning {\tt MAX\_LONG} effectively disables the introspection interceptors.
This is useful, since it allows execution without recompilation of the code even when no tool is used that could determine the bounds of an object.

\section{Introspection in Tools}
\label{sec:tools}
We implemented {\tt \_size\_right()} by exposing existing bounds information in three tools, namely LLVM's AddressSanitizer~\cite{asan}, SoftBound~\cite{softbound}, and GCC's Intel MPX-based Pointer Bounds Checker instrumentation.
SoftBound and LLVM's AddressSanitizer (ASan) are both software-based approaches.
SoftBound provides access to bounds information in constant time, and is therefore a favorable candidate for implementing introspection.
ASan's representation of metadata is suboptimal for implementing introspection, because it does not explicitly maintain bounds information and finding the end of an object takes linear time.
By implementing introspection in ASan, we wanted to determine a worst-case overhead for implementing introspection in existing tools.
Intel MPX instrumentation allowed us to additionally evaluate a hardware-based approach.

\reduceparspace{}
\paragraph{SoftBound.}
SoftBound is a bounds checker that has also been enhanced by a mechanism (called CETS) to find temporal memory errors~\cite{cets}.
It tracks base and bounds information for every pointer as separate metadata.
To propagate this metadata across call sites, SoftBound adds additional base and bounds metadata to pointer arguments of functions.
To implement {\tt \_size\_right()}, we return the right bounds of a pointee by subtracting its base address from its bounds, which are associated with the pointer.
For all SoftBound experiments, we used the latest stable version 3.8.0, which is distributed together with CETS.

\reduceparspace{}
\paragraph{LLVM's AddressSanitizer.}
ASan is one of the most widely used bug-finding tools for C/C++ programs; it allows memory errors such as buffer overflows and use-after-free errors to be found by instrumenting the program during compile time.
Its implementation is based on \emph{shadow memory}~\cite{shadowmemory}, where a memory cell allocated by the program has a corresponding shadow memory cell that stores meta-information about the original allocation.
To detect buffer overflows, ASan allocates space between allocations and marks the corresponding shadow memory as \emph{redzones}; if a dereferenced pointer points to such a redzone, ASan detects the overflow and aborts the program.
Shadow memory is not a favorable representation of metadata for introspection, since bounds information cannot be accessed in constant time.
We implemented {\tt \_size\_right()} in linear time by iterating over the current buffer until its associated shadow memory indicates that a redzone has been reached.
For all LLVM and ASan experiments, we used the development branch of LLVM version 6.0.0 based on commit 1d871d6 in compiler-rt.

\reduceparspace{}
\paragraph{Intel MPX.}
Intel MPX is an instruction set extension that adds instructions for creating, maintaining, and checking bounds information.
Although its performance overhead is relatively high~\cite{mpxexplained}, providing buffer overflow protection at the hardware level is a promising research direction~\cite{cheri}.
To use Intel MPX, we relied on GCC's Pointer Bounds Checker instrumentation, which employs Intel MPX to verify bounds.
Similarly to SoftBound's implementation, we implemented {\tt \_size\_right()} by querying the upper bounds (using a GCC builtin function) and subtracted the pointer address from it.
For all experiments, we used GCC version 7.2.0.

\reduceparspace{}
\paragraph{Using libc.}
To use our introspection-based libc extensions, we redefined the names of the libc functions by means of preprocessor macros.
While this required recompilation of the target application, it allowed the tools to also instrument our introspection-based libc functions and did not require us to maintain bounds information, as libc calls from our interceptors invoked the tools' interceptors.
Note that our approach could be extended by using the dynamic loader to load the interceptors to retain binary compatibility (e.g., using the {\tt LD\_PRELOAD} mechanism); however, redefining the function names was less invasive.

\section{CVE Case Study}
To demonstrate the applicability of our approach in real-world projects, we considered recent (i.e., less than one year old) buffer overflows in widely-used software such as Dnsmasq, Libxml2, and GraphicsMagick.
We selected the first libc-related bugs that we found in the CVE database for which an executable exploit existed.
For each buffer overflow, we evaluated whether our introspection-based approach could mitigate the error and whether execution could successfully continue.
Our approach prevented four out of five buffer overflows while successfully continuing execution; in one case, execution was aborted due to a subsequent buffer overflow in user-level code.
Note that the unmodified tools also detected those buffer overflows; however, they aborted the program instead of mitigating the error and continuing execution.
Since we performed this case study on complex real-world applications, and because SoftBound is a research prototype, we could not successfully execute any of these applications with it.
The unmodified SoftBound version was also unable to execute them.\footnote{https://github.com/santoshn/softboundcets-3.8.0/issues/$x \in \{5, 6, 7, 8\}$}
However, we extracted the functions in which the errors occurred, which SoftBound could execute, and created a driver to trigger the bug.
\reduceparspace{}
\paragraph{Dnsmasq.}
Dnsmasq is a lightweight DHCP server and caching DNS server which is used in many home routers.\footnote{\url{http://www.thekelleys.org.uk/dnsmasq/doc.html}}
In versions prior to 2.78, a bug existed that could cause a stack-based buffer overflow that allowed attackers to execute arbitrary code or to cause denial of service by crafting a DHCPv6 request with a wrong size (see CVE-2017-14493).\pagebreak{}
It occurred in {\tt memcpy()}, to which an incorrect size argument was passed:
{
\small
\begin{verbatim}
state->mac_len = opt6_len(opt) - 2;
memcpy(&state->mac[0], opt6_ptr(opt, 2), state->mac_len);
\end{verbatim}
}

\noindent A similar bug could be exploited for denial of service attacks (see CVE-2017-14496).
It occurred in {\tt memset()} and was triggered by an integer overflow: 
{
\small
\begin{verbatim}
/* Clear buffer beyond request to avoid risk of information disclosure. */
memset(((char *)header) + qlen, 0, (limit - ((char *)header)) - qlen);
\end{verbatim}
}
\noindent When using our introspection interceptors, all tools continued execution by copying or setting up to as many bytes as the destination buffer could hold.
The server stayed fully functional.

\reduceparspace{}
\paragraph{Libxml2.}
Libxml2 is a widely used open-source XML parsing library.\footnote{\url{http://xmlsoft.org/}}
For versions up to 2.9.4, a vulnerability in the {\tt xmlSnprintfElementContent()} function enabled attackers to crash the application through a buffer overflow (see CVE-2017-9047).
It was caused by an incorrect length validation (at another code location) followed by {\tt strcat()}:

{
\begin{small}
\begin{verbatim}
if (content->name != NULL)
  strcat(buf, (char *) content->name);
\end{verbatim}
\end{small}
}
\noindent The introspection interceptor for {\tt strcat()} mitigated the buffer overflow by restricting the length of the concatenated string in all tools.
The application continued execution and printed the truncated string as part of an error message.
Although the error message was truncated, the output appeared reasonable from the user's point of view.

\reduceparspace{}
\paragraph{GraphicsMagick.}
GraphicsMagick is a widely used image processing tool.\footnote{\url{http://www.graphicsmagick.org/}}
In version 1.3.26, its {\tt DescribeImage()} function allowed attackers to overflow and corrupt the heap to execute arbitrary code or to cause denial-of-service attacks (see CVE-2017-16352). 
As shown below, the size argument in the call to {\tt strncpy()} did not limit the number of copied bytes to the size of the buffer; instead, the number was calculated by the length of the directory name (which was determined by searching for the newline or {\tt NUL}).
Consequently, an overly long directory name could be used to cause an overflow:
{
\small
\begin{verbatim}
for (p=image->directory; *p != '\0'; p++) {
  q=p;
  while ((*q != '\n') && (*q != '\0'))
    q++;
  (void) strncpy(image_info->filename,p,q-p);
  image_info->filename[q-p]='\0';
  p=q;
  // ...
}
\end{verbatim}
}
\noindent The introspection interceptor for {\tt strncpy()} successfully restricted the length of the copied string to the length of the destination buffer {\tt image\_info->filename}.
However, in the line after the call to {\tt strncpy()}, the program attempted to write a {\tt NUL} character to the end of the string, which then caused an out-of-bounds access in the user application.
The introspection approach does not protect against buffer overflows that happen in code that does not use introspection; however, we intend introspection to be used together with a bounds-checking tool, which is expected to abort execution for unhandled errors and thus prevent incorrect execution.
In fact, all introspection-instrumented tools prevented this buffer overflow by aborting execution.

\reduceparspace{}
\paragraph{LightFTP.}
LightFTP is a small FTP server.\footnote{\url{https://github.com/hfiref0x/LightFTP/}}
A logging function {\tt writelogentry()} in version 1.1 of LightFTP was vulnerable to a buffer overflow that allowed denial of service or remote code execution (see CVE-2017-1000218).
As shown below, the program added log entries to a buffer with a hard-coded size; as the log entries depended on user input that was restricted by another, larger constant, a buffer overflow could be triggered: 

{
\small
\begin{verbatim}
char _text[512];
// ...
if (logtext1)
    strcat(_text, logtext1);
if (logtext2)
    strcat(_text, logtext2);
strcat(_text, CRLF);
\end{verbatim}
}

\noindent The introspection interceptor for {\tt strcat()} mitigated the error without crashing the FTP server.
Note that our mitigation truncated the log entry, but allowed subsequent requests to be handled successfully.

\section{Performance Evaluation}
\begin{figure}[bt!]
    \centering
    \includegraphics[width=11cm]{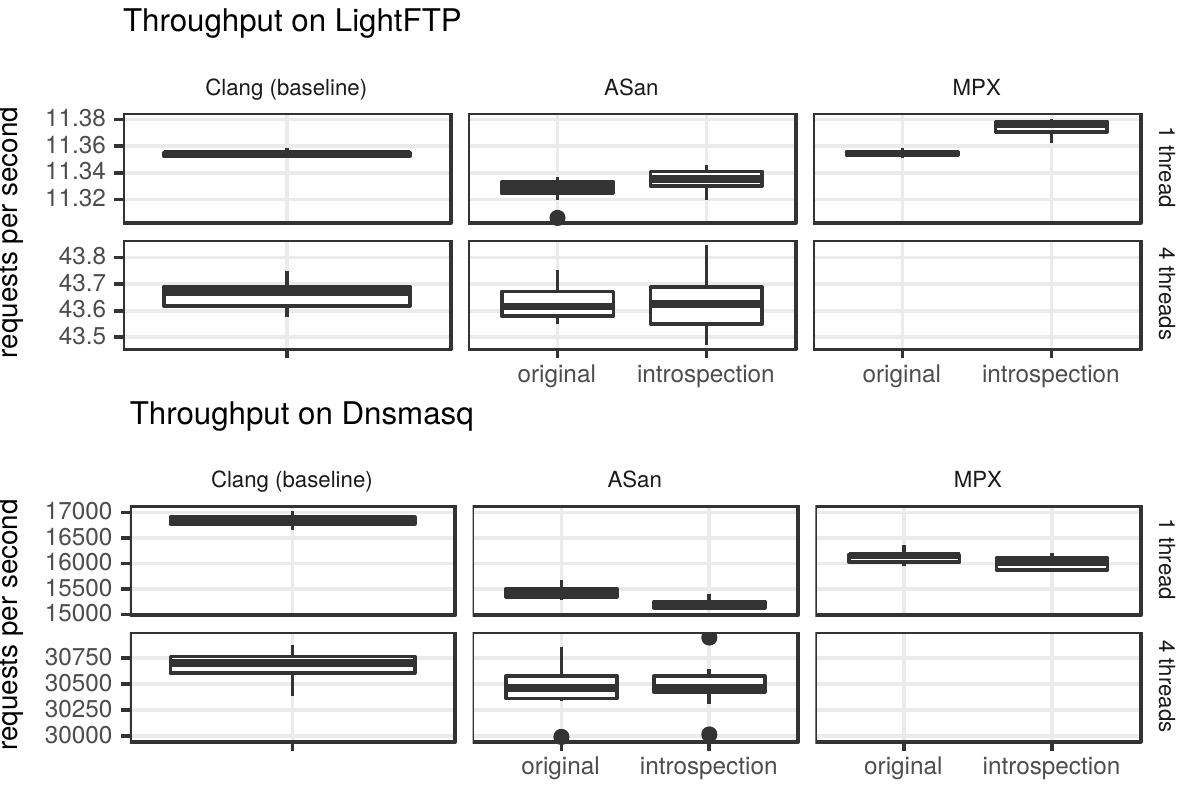}
    \caption{Throughput on LightFTP and Dnsmasq.}
    \label{plot:throughput}
\end{figure}

To determine the performance of the introspection-based interceptors, we used LightFTP and Dnsmasq, which are the servers we also investigated in our CVE case study.
We selected them for their high attack surface and because they are expected to be highly available.
We evaluated the performance of ASan and Intel MPX both with and without the introspection interceptors; SoftBound failed to execute the servers, as explained above.
Further, to establish a baseline, we measured the performance of C programs compiled with the Clang compiler~\cite{llvm} without using any bug-mitigation mechanisms.
For all systems, we turned on compiler optimizations by using the {\tt -O3} flag.
We measured the throughput by means of the load-testing tool JMeter version 3.3.
We configured JMeter to use 4 threads, each of which each sent 250 requests to simulate multiple concurrent users using the  built-in FTP sampler and the UDP Protocol Support plugin.
As the Intel MPX instructions are not thread-safe~\cite{mpxexplained}, we also evaluated all tools using only 1 thread.
We performed each measurement 10 times to account for variability.
Our setup consisted of a quad-core Intel Core i7-6700HQ CPU at 2.60GHz on Ubuntu version 17.10 (with kernel 4.13.0-32-generic) with 16 GB of memory.

Figure~\ref{plot:throughput} shows boxplots of the results for LightFTP and Dnsmasq.
On LightFTP, the performance overhead for using introspection was below 1\% for ASan; MPX was even slightly faster when introspection was used.
On Dnsmasq, employing introspection caused a slowdown of around 1\% when using only one thread for both ASan and MPX.
The performance difference to the baseline was negligible on LightFTP, and up to 11\% on Dnsmasq (between Clang and ASan with introspection), which suggests that the applications' performance was dominated by factors other than instrumentation cost (e.g., networking overhead).
Thus, our measurements cannot be generalized to CPU-bound benchmarks.

To quantify the overhead on CPU-bound benchmarks, we also evaluated the approaches on the SPEC2006 INT benchmarks, which consist of 12 benchmarks.
We excluded all C++ benchmarks (471.omnetpp, 473.astar, and 483.xalancbmk), which we expected to make little use of C functions and thus of our interceptors.
Further, we excluded all benchmarks in which the tools detected memory safety errors (400.perlbench and 403.gcc).
ASan detected memory leaks in two benchmarks (445.gobmk and 464.h264ref), and since we investigated only buffer overflows in this work, we disabled memory leak detection to also run them.
SoftBound in its original and introspection versions detected memory safety errors in all but one benchmark (458.sjeng), which were presumably false positives.
MPX had an additional known false positive~\cite{mpxexplained} in one benchmark (429.mcf), so we excluded this benchmark for MPX.

\begin{figure}[bt!]
    \includegraphics[width=\columnwidth]{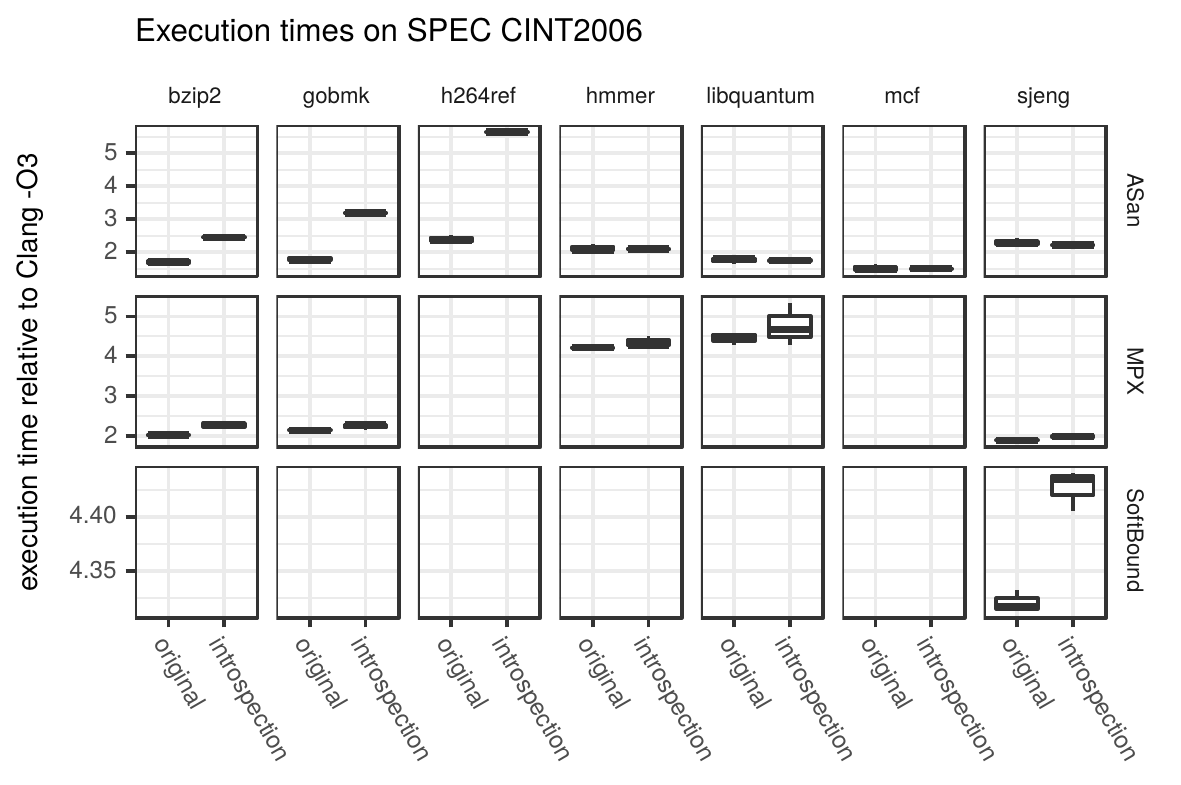}
    \caption{Execution times on the SpecInt2006 benchmarks.}
    \label{plot:spec}
\end{figure}

Figure~\ref{plot:spec} shows the execution times of the SPECInt2006 benchmarks relative to Clang -O3 as a baseline.
On four of the seven benchmarks (429.mcf, 456.hmmer, 458.sjeng, 462.libquantum), the performance overhead was negligible because no interceptors were executed in code that contributed to the overall run-time performance of the respective benchmark.
For SoftBound, the introspection overhead was 3\% on the only benchmark that it could execute.
Using introspection with ASan resulted in higher overheads, namely 140\% on h264ref, 43\% on bzip2, and 81\% on gobmk.
For MPX, the performance overhead of introspection was relatively low, with maximum overheads of 13\% on bzip2 and 6\% on gobmk.

We also executed micro-benchmarks, measuring the direct overhead of interceptors.
For example, we evaluated the performance of the {\tt strlen()} interceptor, which directly relies on {\tt \_size\_right()} to call the safer {\tt strnlen()} function. 
For SoftBound, the overhead was not measurable.
For Intel MPX, the overhead was 2$\times$ for strings with a length of 10; for longer strings (e.g., a length of 1000) the overhead was not measurable.
The overhead for ASan was the highest, as our {\tt size\_right()} implementation has to traverse the shadow memory, which depends linearly on the length of the string.
Its overhead varied between 2$\times$ and 10$\times$ with different string lengths.

\section{Discussion}
\paragraph{Availability.} We have demonstrated that our introspection-based libc interceptors are an effective means of mitigating the effects of buffer overflows.
Our main idea is to use run-time information that is tracked by existing tools to prevent buffer overflows and to increase the availability of applications.
Using the introspection-based interceptors is useful only in production, because during development and testing it would be preferable to abort execution so that the programmer can fix bugs that cause errors.

\paragraph{Complementarity.}
We have designed our approach to complement existing approaches for handling buffer overflows.
Our idea is that, for important functions, programmers can implement custom semantics that mitigate the effects of buffer overflows.
For buffer overflows in other functions or in user-level code, existing memory tools would continue to detect out-of-bounds accesses and would abort execution in the case of an error.
Alternatively, the interceptors could also be used with the original failure-oblivious computing approach as a fallback for functions that are not guarded by introspection checks.

\reduceparspace{}
\paragraph{Performance.}
The overhead of introspection and our interceptors depends mainly on how efficiently a tool tracks bounds information.
Our evaluation on servers suggests that the overhead of introspection is often small compared to the cost of network communication, making introspection especially applicable for servers.
Our evaluation on the CPU-bound SPEC benchmarks also seems to suggest that libc functions are typically not part of the code that significantly contributes to the overall performance of a program.
While the MPX-based introspection overhead was low on all benchmarks, only the ASan-based implementation caused larger overheads on three benchmarks.
Overall, introspection-based libc functions are feasible with a low overhead for approaches that maintain explicit bounds information (e.g., Intel MPX or SoftBound), but result in higher overheads for approaches in which bounds information must be computed (e.g., in ASan).
Furthermore, our implementation could be made more efficient by using introspection directly in the libc functions.

\reduceparspace{}
\paragraph{Implementation.} We have demonstrated implementations of the {\tt \_size\_right()} function for three popular bug-finding and bug-mitigation approaches and believe that implementing this function in many others (e.g.,~libcrunch~\cite{Kell2015,libcrunch}) is also straightforward.
Some tools cannot give precise estimates for all pointers, which makes our approach less effective.
For example, binary-instrumentation tools such as Valgrind~\cite{valgrind} and Dr.~Memory~\cite{drmemory} cannot reliably determine the size of buffers located on the stack.
Other approaches track run-time information only for specific types of allocations (e.g., stack buffers~\cite{libsafe}).
Furthermore, some tools give rough estimates in general or round up allocation sizes~\cite{baggyboundschecking,libsafe,lowfatheap}; for example, after evaluating our approach with low-fat pointer checking~\cite{lowfatheap,Duck2017}, we found that rounding up allocation sizes alone mitigated several of the buffer overflows that we investigated.\footnote{EffectiveSan~\cite{Duck2018}, an extension of the low-fat pointer approach, provides accurate bounds but has not been released to the public as of June 2018.}
Note that conservative estimates (e.g., the maximum integer value if no information is available) ensure correct execution, but might result in undetected errors.

\section{Related Work}

\paragraph{Failure-oblivious computing.}
Rinard et al. coined the term \emph{failure-oblivious computing}, where illegal read accesses yield predefined values and out-of-bounds write accesses are ignored~\cite{failureoblivious}.
An extension of this work are \emph{boundless memory blocks}, where out-of-bounds writes store the value in a hash map that can be returned for out-of-bounds reads to that address~\cite{boundlessmemoryallocations,sgxbounds,boundlessmemoryblock}.
Furthermore, Long et al. extended failure-oblivious computing by also covering divide-by-zero errors and {\tt NULL}-pointer dereferences~\cite{Long2014}.
In contrast to these approaches, introspection enables programmers to handle out-of-bounds accesses by taking into account the semantics of a function. 
However, the drawback of our approach is that library developers must implement these checks manually.

\reduceparspace{}
\paragraph{Failure-oblivious computing models.}
Durieux et al. studied failure-oblivious computing behaviors~\cite{Durieux2018}.
Their findings suggest that for many failures, multiple alternative strategies exist that can mitigate the error.
For example, to mitigate a {\tt NULL}-pointer dereference the access could be ignored, but the pointer could also be initialized with the address of a newly-created or existing object.

\reduceparspace{}
\paragraph{Monitored execution.}
Sidiroglou et al. devised a system that monitors an application for failures such as buffer overflows~\cite{immunesystem}.
If a fault occurs, the current function is aborted and---based on heuristics---an appropriate value is returned.
In order to avoid crashes because a pointer returns {\tt NULL}, the heuristics take into account whether the parent function dereferences the pointer thereafter.
While this approach takes into account the context of the fault, it lacks the ability of our introspection approach to benefit from programmer knowledge.

\paragraph{Libsafe.}
Libsafe replaces libc functions with enhanced versions that prevent buffer overflows from going beyond the stack frame~\cite{libsafe}.
It achieves this by traversing frames to determine their bounds and aborting the program if the bounds are exceeded.
While we tried implementing the introspection function using the traversal logic, we found that it is based on assumptions such as the location of the stack, which no longer hold with modern mitigation techniques such as address space layout randomization.
Additionally, libsafe does not handle out-of-bounds reads well, for which our approach, in contrast, can compute meaningful results, for example, by letting {\tt strlen()} return the length of the buffer underlying the string if it is unterminated.

\reduceparspace{}
\section{Conclusion}
In this paper, we have presented how implementation of an introspection function that returns the length of an object can be used to implement failure-oblivious computing mechanisms.
We have also shown that such a mechanism is useful in mitigating real-world errors and that the performance overhead when implemented in approaches such as Intel MPX is negligible.
For reproducibility and to facilitate further research, we distribute all artifacts and experimentation scripts at
{\tt \url{https://github.com/introspection-libc/main}}.

\bibliographystyle{splncs04}
\bibliography{nss}
\end{document}